\documentclass[preprint]{ptptex}

\usepackage{graphicx}
\usepackage{wrapft}
\usepackage{multirow}

\def \ketv #1>{\mbox{$|{#1}\rangle$}} 
\def \brav #1|{\mbox{$\langle {#1}|$}}
\def \mate<#1|#2|#3>{\mbox{$\langle {#1}|\,{#2}\,|{#3}\rangle$}}

\title{
Baryon-Baryon Interactions in the Flavor SU(3) Limit \\ 
from Full QCD Simulations on the Lattice
}

\author{
Takashi \textsc{Inoue}$^{1,}$\footnote{E-mail: inoue.takashi@nihon-u.ac.jp},
Noriyoshi \textsc{Ishii}$^{2}$,
Sinya \textsc{Aoki}$^{3,4}$,
Takumi \textsc{Doi}$^{3}$,\\
Tetsuo \textsc{Hatsuda}$^{2}$,
Yoichi \textsc{Ikeda}$^{5}$,
Keiko \textsc{Murano}$^{6}$,
Hidekatsu \textsc{Nemura}$^{7}$,\\
Kenji \textsc{Sasaki}$^{3}$ (HAL QCD collaboration)\\
\bigskip
\includegraphics[width=0.3\textwidth]{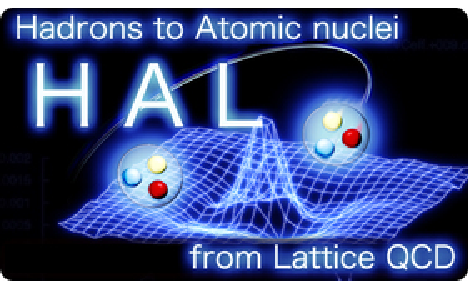}
}

\inst{
$^1$Nihon University, College of Bioresouce Sciences, Kanagawa 252-0880, Japan\\
$^2$Department of Physics, University of Tokyo, Tokyo 113-0033, Japan\\
$^3$Graduate School of Pure and Applied Sciences, University of Tsukuba,\\
    Ibaraki 305-8571, Japan\\
$^4$Center for Computational Sciences, University of Tsukuba, Tsukuba, Ibaraki 305-8577, Japan\\
$^5$Nishina Center for Accelerator-Based Science, Institute for Physical and Chemical Research (RIKEN), Saitama 351-0198, Japan\\
$^6$High Energy Accelerator Research Organization (KEK), Ibaraki 305-080, Japan\\
$^7$Department of Physics, Tohoku University, Miyagi 980-8578, Japan
}

\abst{
We investigate baryon-baryon (BB) interactions 
in the 3-flavor full QCD  simulations with degenerate quark masses for all flavors. 
The BB potentials in the orbital S-wave are extracted from the Nambu-Bethe-Salpeter 
wave functions measured on the lattice.
We observe strong flavor-spin dependences of the BB potentials at short distances.
In particular, a strong repulsive core exists in the
flavor-octet and spin-singlet channel
 (the ${\bf 8}_s$ representation),
while an attractive core appears in the flavor singlet channel (the  ${\bf 1}$ representation).
 We discuss a relation of such flavor-spin dependence with the Pauli exclusion principle 
 in the quark level.  Possible existence of an $H$-dibaryon resonance above the $\Lambda\Lambda$ 
threshold is also discussed.
}

\begin{document}

\maketitle

\section{Introduction}
The generalized nuclear force, which includes not only the nucleon-nucleon (NN) interaction 
but also hyperon-nucleon (YN) and hyperon-hyperon (YY) interactions,
has been one of the central topics in hadron and nuclear physics.
Experimental studies on the ordinary and hyper nuclei \cite{Hashimoto:2006aw} 
as well as the observational studies of the neutron stars and supernova explosions 
\cite{SchaffnerBielich:2010am} are
intimately related to the physics of the generalized nuclear force.

As far as the  NN interaction is concerned, significant number of
scattering data have been accumulated. 
Furthermore, the phenomenological NN potential is
one of the useful means to parametrize the phase shift data below
the meson production threshold:
 Indeed,
the S-wave NN phase shift can be described well by a potential 
 in the coordinate space
with a repulsive core at short distance and  an attractive well
 at intermediate and long distances \cite{NF_I-II,Machleidt:2001rw,Epelbaum:2008ga}. 

In contrast to the NN interaction, properties of the YN and YY 
interactions are poorly known even today
due to the lack of high precision  scattering data for hyperons.
The approximate flavor SU(3) symmetry in the hadronic level
does not fully constrain the hyperon interactions since
there exists six independent flavor-channels for the scattering of octet baryons.
Theoretical attempts to extract the YN and YY interactions have been made
on the basis of the constituent quark models\cite{Oka:2000wj}
or the one-boson-exchange models\cite{Rijken:2010zz},
but there still exists a large uncertainty for the determination of the generalized nuclear force.
To improve this situation, a model independent investigation based on QCD is 
called for
\cite{Aoki:2009yy}. 

Recently a new method to extract the baryon-baryon potentials in the 
coordinate space from lattice QCD simulations
has been proposed and applied to the NN system\cite{Ishii:2006ec,Aoki:2009ji}, 
and $\Xi N$ and $\Lambda N$ systems\cite{Nemura:2008sp,Nemura:2009lat}.
In this method, energy independent non-local potentials are defined
from the Nambu-Bethe-Salpeter (NBS) amplitude through the Schr\"odinger equation.
In the above studies, it has been shown that the 
repulsive core at short distance appears in all NN, $\Xi N$ and $\Lambda N$ systems,
while the height and range of the repulsion depend on flavor and spin combinations
of two baryons.
 
The purpose of this paper is to make a systematic study
on the short range baryon-baryon (BB) interactions with full QCD simulations on the lattice.
(A preliminary account is given in Ref.\cite{Inoue:2009ce}).
This problem is also closely related to possible dibaryon states
such as the flavor-singlet six-quark system $H$ in phenomenological 
quark models\cite{Jaffe:1976yi,Sakai:1999qm}.
To capture  essential features of the short range force
without contamination from the quark mass difference,
we consider the flavor SU(3) limit where all $u$, $d$, and  $s$ quarks have a common finite mass.
This allows us to extract BB 
potentials for irreducible flavor multiplets, from  recombinations of which
the potentials between asymptotic baryon states
are obtained with suitable Clebsch-Gordan (CG) coefficients. 

This paper is organized as follows.
In Section 2, we explain a classification of BB states in the flavor SU(3) limit and
a method to extract BB potentials in lattice QCD.
In Section 3, we give a brief summary of our numerical simulations.
In Section 4, we present the resulting BB potentials and discuss their implications. 
The last section is devoted to summary and outlook.

\section{\label{sec:su3}Baryon-baryon potentials in the flavor SU(3) limit}

In the flavor SU(3) limit, the ground state baryons belong to
the flavor-octet with spin 1/2 or the flavor-decuplet with spin 3/2.
In this paper,   we focus on the octet baryons, for which
two baryon states with a given angular momentum
are labeled by the irreducible flavor multiplets as
\begin{equation}
 {\bf 8} \otimes {\bf 8} 
 = \underbrace{{\bf 27} \oplus {\bf 8}_s \oplus {\bf 1}}_{\mbox{symmetric}} ~ 
  \oplus \underbrace{{\bf 10}^* \oplus {\bf 10} \oplus {\bf 8}_a}_{\mbox{anti-symmetric}} \ . 
\end{equation}
Here ``symmetric" and ``anti-symmetric" stand for the symmetry under the
flavor exchange of two baryons.
For the system in the orbital S-wave, the Pauli principle between two baryons imposes 
${\bf 27}$, ${\bf 8}_s$ and ${\bf 1}$ to be spin singlet  ($^1S_0$) while 
${\bf 10}^*$, ${\bf 10}$ and ${\bf 8}_a$ to be spin triplet ($^3S_1$). 
Since there are no mixings among different multiplets in the SU(3) limit, 
one can define the corresponding potentials as
 \begin{eqnarray}
^1S_0 \ &:& \  V^{({\bf 27})}(r), \ V^{({\bf 8}_s)}(r), \ V^{({\bf 1})}(r), 
\\ 
^3S_1 \ &:& \ V^{({\bf 10}^*)}(r), \ V^{({\bf 10})}(r), \ V^{({\bf 8}_a)}(r) ~.
\end{eqnarray}
Potentials among octet baryons, both the diagonal part ($B_1B_2 \rightarrow B_1 B_2)$ and  
the off-diagonal part ($B_1B_2 \rightarrow B_3 B_4$), are obtained by  suitable combinations
of $V^{(\alpha)}(r)$ with $\alpha={\bf 27},{\bf 8}_s,{\bf 1},{\bf10}^*,{\bf 10},{\bf 8}_a$.

According to Ref. \cite{Ishii:2006ec,Aoki:2009ji} , the above potentials can be defined
from the NBS wave function $\phi^{(\alpha)}(\vec r)$ 
through the Schr\"odinger equation as
\begin{equation}
   \left[ \frac{\nabla^2}{2 \mu} + E^{(\alpha)} \right] \phi^{(\alpha)}(\vec r) 
 = \int\!\!d^3{\vec r}' \, U^{(\alpha)}(\vec r, {\vec r}') \, 
 \phi^{(\alpha)}({\vec r}'),
 \label{eq:shroedinger}
\end{equation}
with $ E^{(\alpha)} = {k^2}/{(2\mu)}$ and $k^2 = {W^2}/{4} - {M^2}/{4}$.
Here $\mu=m/2$, $M=2m$ and $W$ are the reduced mass, total mass and the 
total energy in the center of mass frame of the two baryon system
 with $m$ being the octet baryon mass in the SU(3) limit.
Note that  $U^{(\alpha)}(\vec r, \vec r')$ is non-local but energy-independent
 and may be written as
  $U^{(\alpha)}(\vec r, \vec r')=V^{(\alpha)}(\vec r, \nabla) \delta^3 ({\vec r} - {\vec r}')$
  \cite{Aoki:2009ji}.
At the leading-order of the derivative expansion of $V^{(\alpha)}(\vec r, \nabla)$
\cite{Aoki:2009ji,Murano:2010tc},
we obtain an effective central potential\footnote{
The ``effective" central potential is obtained from the NBS wave function 
without making a decomposition into central and tensor potentials
in the leading order of the derivative expansion \cite{Ishii:2006ec,Aoki:2009ji}.
In the $^1S_0$ state, the ``effective" central potential reduces to the central potential
due to the absence of the tensor component.
}  
\begin{equation}
 V^{(\alpha)}(r) = \frac{1}{2\mu}
 \frac{\nabla^2 \phi^{(\alpha)}(\vec r)}{\phi^{(\alpha)}(\vec r)} + E^{(\alpha)} ~.
 \label{eqn:vr}
\end{equation}

The NBS wave function $\phi^{(\alpha)}$ is defined by
\begin{equation}
\phi^{(\alpha)}(\vec r)e^{-Wt}
= \langle 0 \vert (BB)^{(\alpha)}(t,\vec r) \vert B=2, \alpha, W \rangle,
\end{equation}
where 
$(BB)^{(\alpha)} (t,\vec r)=\sum_{i,j,\vec x} C_{ij} B_i(t,\vec x+ \vec r) B_j (t,\vec x)$
is a two-baryon operator with relative distance $\vec r$ in  $\alpha$-plet and 
$\vert B=2, \alpha, W \rangle $ is a QCD eigenstate with the total energy $W$
and the baryon number 2 in the $\alpha$-representation. 
Here $B_i$ is an one-baryon composite field operator in the octet. 
The relation between two-baryon operators in the flavor basis and baryon basis
are given in Appendix \ref{app:source}.
In the lattice QCD simulations, the above NBS wave function for the smallest $W$
is extracted from the four point function as
\begin{eqnarray}
 G_4^{(\alpha)}(t-t_{0},\vec r) 
 =  \mate<0|(BB)^{(\alpha)} (t,\vec r)\  \overline{(BB)}^{(\alpha)}(t_0)|0> 
 \propto   \phi^{(\alpha)}(\vec r)
 e^{- W(t-t_0)}
\end{eqnarray}
for $t-t_0 \gg 1$, where $\overline{(BB)}^{(\alpha)}(t_0)$ is
a wall source operator at time $t_0$ to create two-baryon states in $\alpha$-plet
while $(BB)^{(\alpha)} (t,\vec r)$ is the sink operator at time $t$ to annihilate two-baryon states.

\section{Numerical simulations}

\begin{table}[t]
\caption{\label{tbl:lattice} Summary of lattice parameters and hadron masses.
The value of lattice spacing $a$ is determined from $\rho$ meson mass at the physical light and strange quark masses.
See the official home-page of CP-PACS and JLQCD Collaborations\cite{CPPACS-JLQCD}.}
 \begin{tabular}{c|c|c|c|c|c|c|c}
   \hline \hline
    lattice  & $\beta$ & ~a [fm] ~& ~L [fm] ~  
             & $\kappa_{uds}$  & ~$m_{\rm ps}$ [MeV]~ & ~ $m_{B}$ [MeV]~ & ~$N_{\mbox{cfg}}$~ \\
   \hline 
   \multirow{2}{*}{$16^3 \times 32$} & ~\multirow{2}{*}{1.83}~
                                     & \multirow{2}{*}{0.121(2)} & \multirow{2}{*}{1.93(3)}
          & ~0.13710~ & 1014(1) & 2026(3) & 700 \\
    & & & & ~0.13760~ & ~835(1) & 1752(3) & 800 \\
   \hline \hline
 \end{tabular}
\end{table}

We employ the gauge configurations
generated by CP-PACS and JLQCD Collaborations with the renormalization group improved Iwasaki gauge action
and the non-perturbatively $O(a)$ improved Wilson quark action\cite{Ishikawa:2007nn}.
These configurations are provided by Japan Lattice Data Grid (JLDG) 
and International Lattice Data Grid (ILDG).
Quark propagators are calculated  for the spatial wall source at $t_0$
with the Dirichlet boundary condition in temporal direction at $t=t_0\pm 16$ mod 32.
In order to enhance the signal,  all 32  time slices  on each configuration are used as $t_0$ of the wall source.
In addition we carry out  an average over forward and backward propagations in time. 

Our lattice parameters are summarized in Table \ref{tbl:lattice}.
Those hopping parameters reside around the physical strange quark mass
region in 2+1 flavor QCD at the same $\beta$\cite{CPPACS-JLQCD}.
Since the typical range of the BB interactions at this quark mass
is about 1 fm or less as shown below, we expect that the 
small lattice size ($L \simeq 2$~fm) would not affect 
the short range part of the BB interactions qualitatively. 

In our calculation, the sink operator is projected to the $A_{1}^{+}$ representation 
of the cubic group,
so that the NBS wave function is assumed to be dominated by the S-wave component. 
We estimate the statistical errors by the jackknife method with a bin size
of seven for $\kappa_{uds}=0.13710$ and  eight  for $\kappa_{uds}=0.13760$.
All of these numerical computations have been carried out 
at KEK supercomputer system, Blue Gene/L and SR11000.

\section{Results and their implications}

\begin{figure}[tp]
 \includegraphics[width=0.49\textwidth]{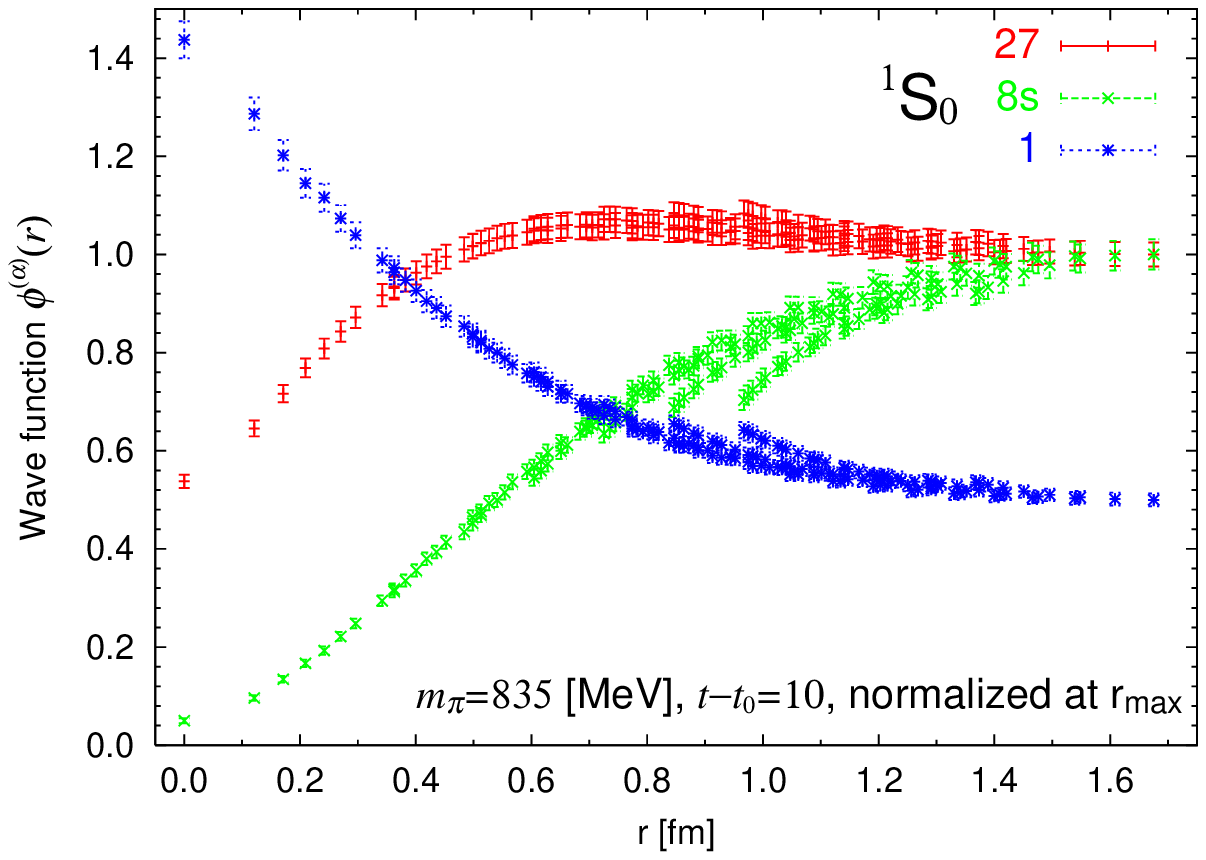}
 \includegraphics[width=0.49\textwidth]{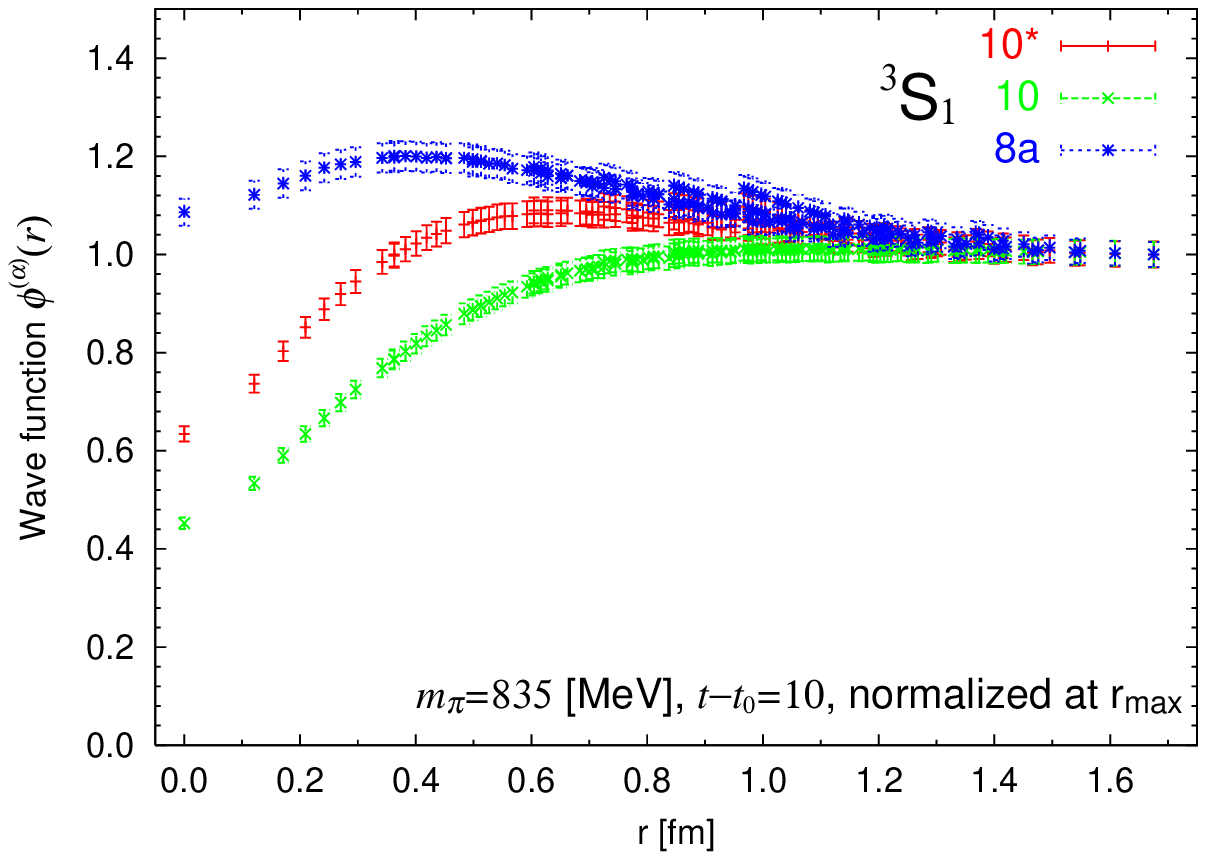}
\caption{\label{fig:wave}
 NBS wave functions  at $m_{\pi}=835$ MeV and $t-t_0=10$, normalized to 1/2 for the singlet channel
 and to 1 for other channels at the maximum distance.}
\end{figure}

\subsection{BB potentials in flavor basis}

Fig. \ref{fig:wave} shows the NBS wave functions as a function
of the relative distance between two baryons 
at  $m_{\pi}=835$ MeV and $t-t_0=10$. 
To draw all data in a same scale, they are normalized to 1/2 for the singlet channel
and to 1 for other channels at the maximum distance.
Since the effective mass of the four point function in each channel shows 
a plateau at  $t-t_0 \ge 10$, we use data at $t-t_0=10$ exclusively
to extract both NBS wave functions and BB potentials throughout this paper. 
The wave functions in Fig. \ref{fig:wave} show characteristic flavor dependence:
In particular, a strong suppression at short distance appears in the ${\bf 8}_s$ channel,
while a strong enhancement appears in the ${\bf 1}$ channel.
Similar results are obtained for $m_{\pi}=1014$ MeV too. 
 
\begin{figure}[tp]
 \includegraphics[width=0.49\textwidth]{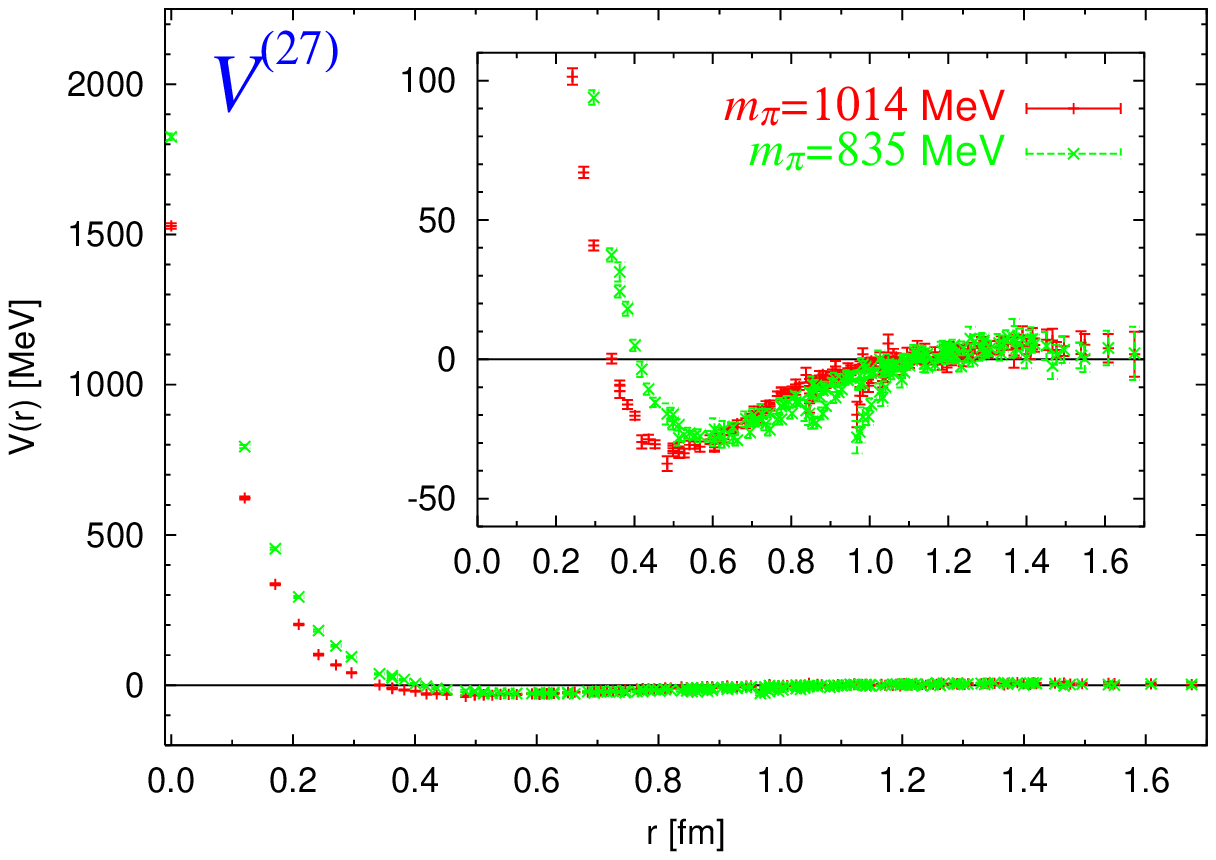}\hfill
 \includegraphics[width=0.49\textwidth]{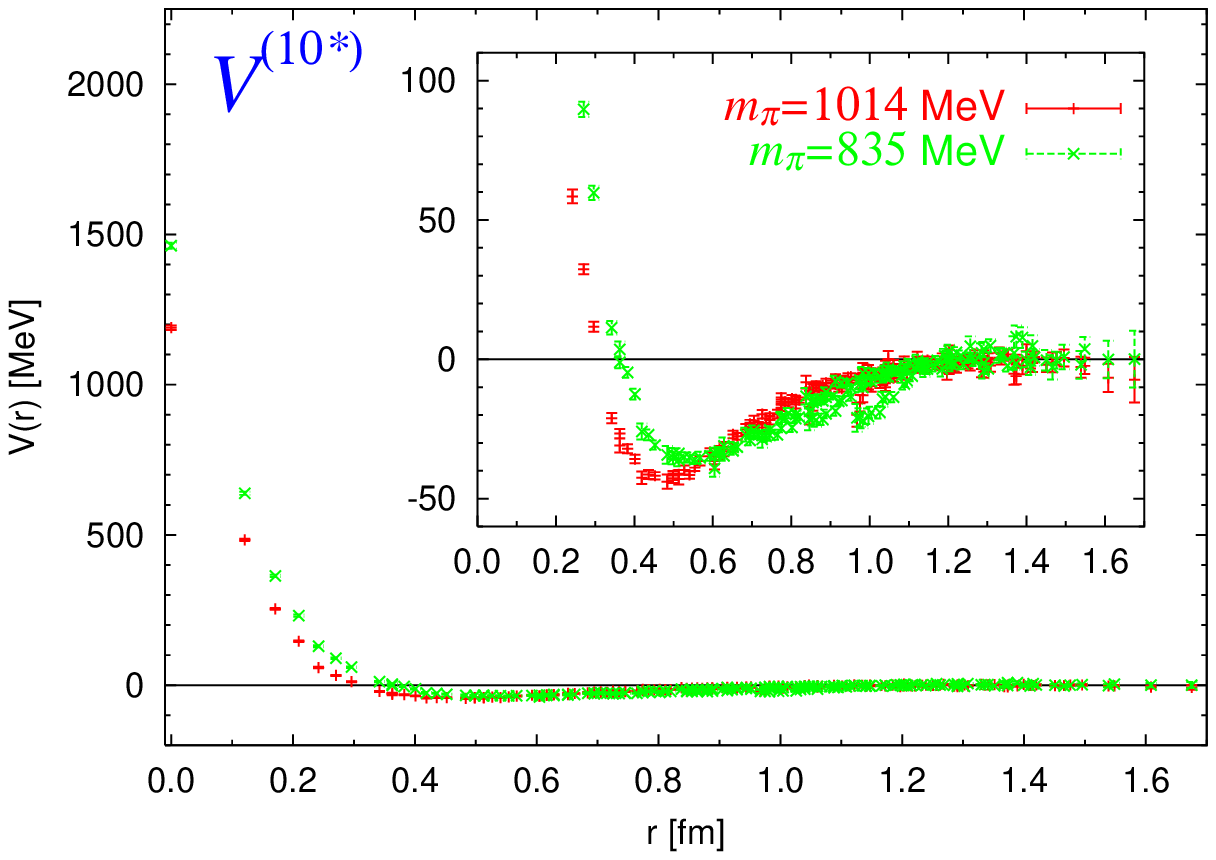}
 \includegraphics[width=0.49\textwidth]{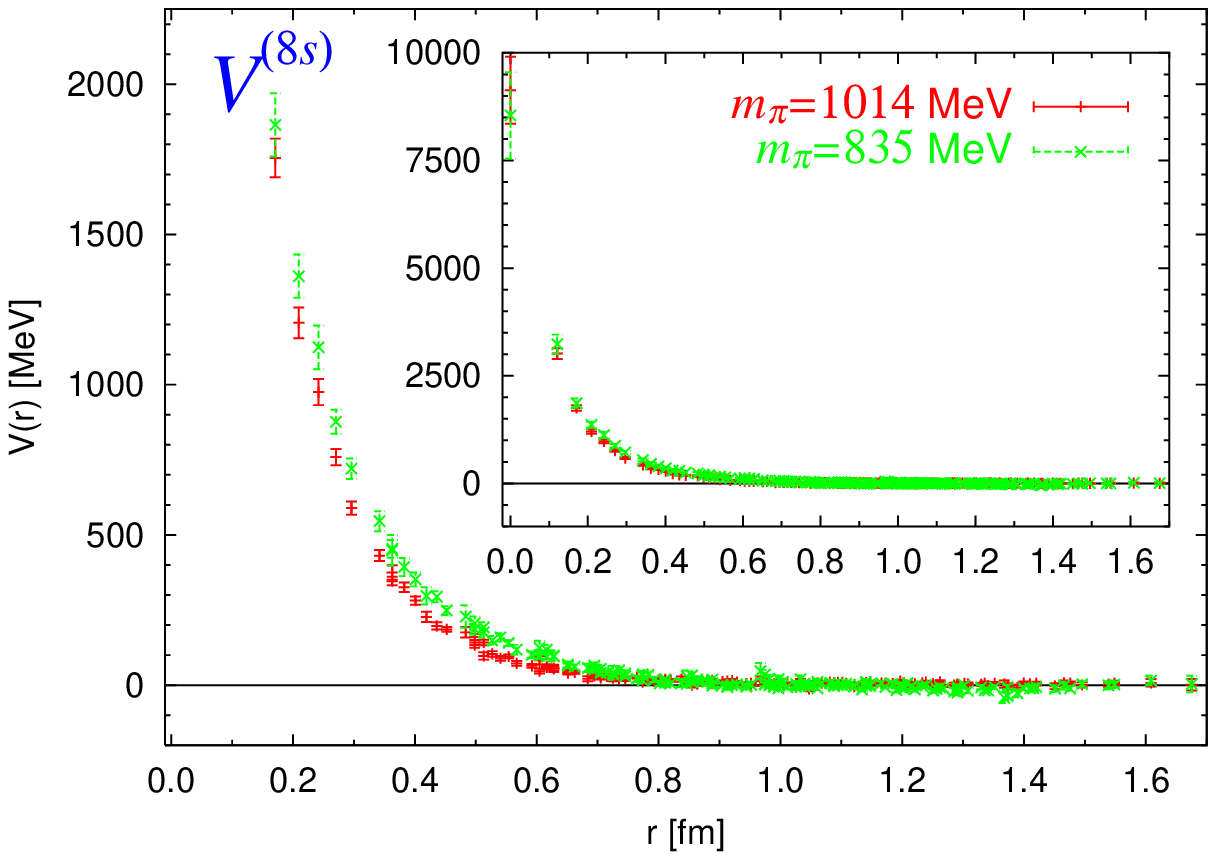}\hfill
 \includegraphics[width=0.49\textwidth]{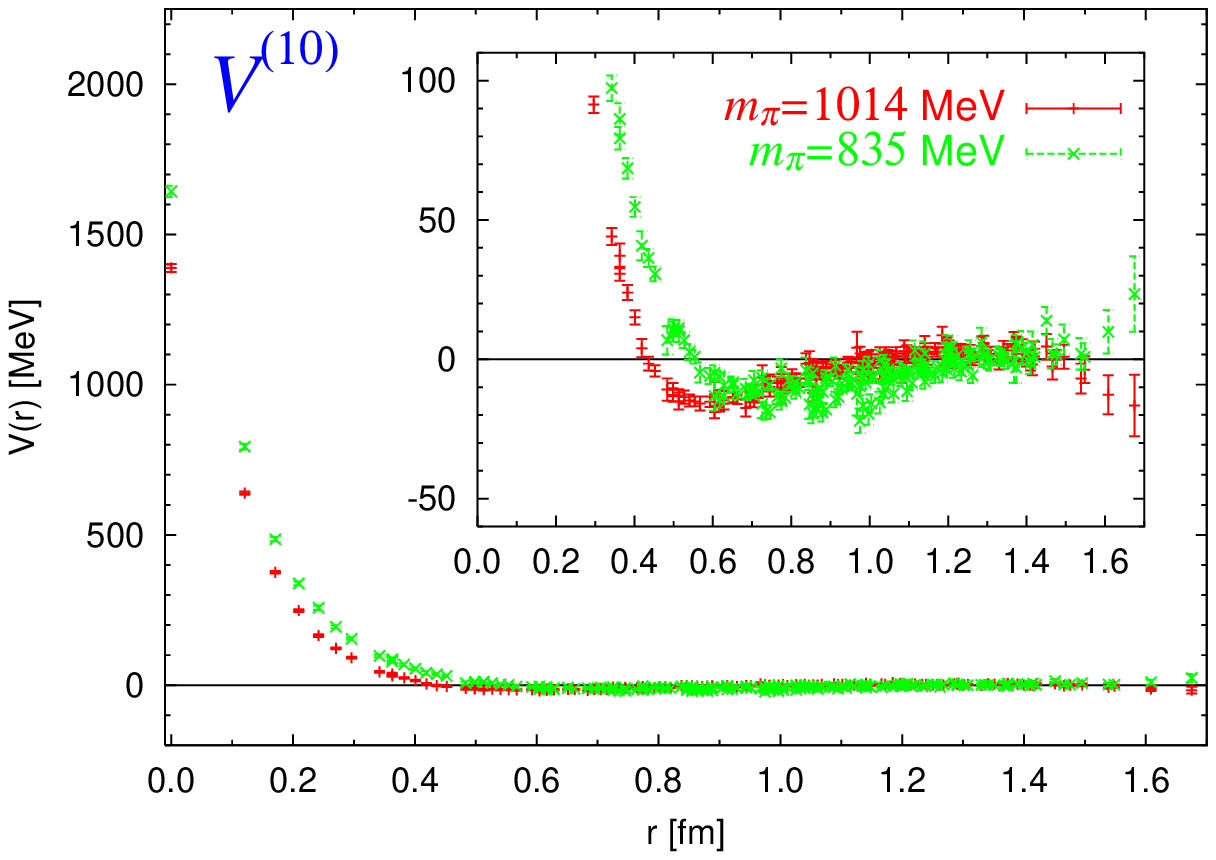}
 \includegraphics[width=0.49\textwidth]{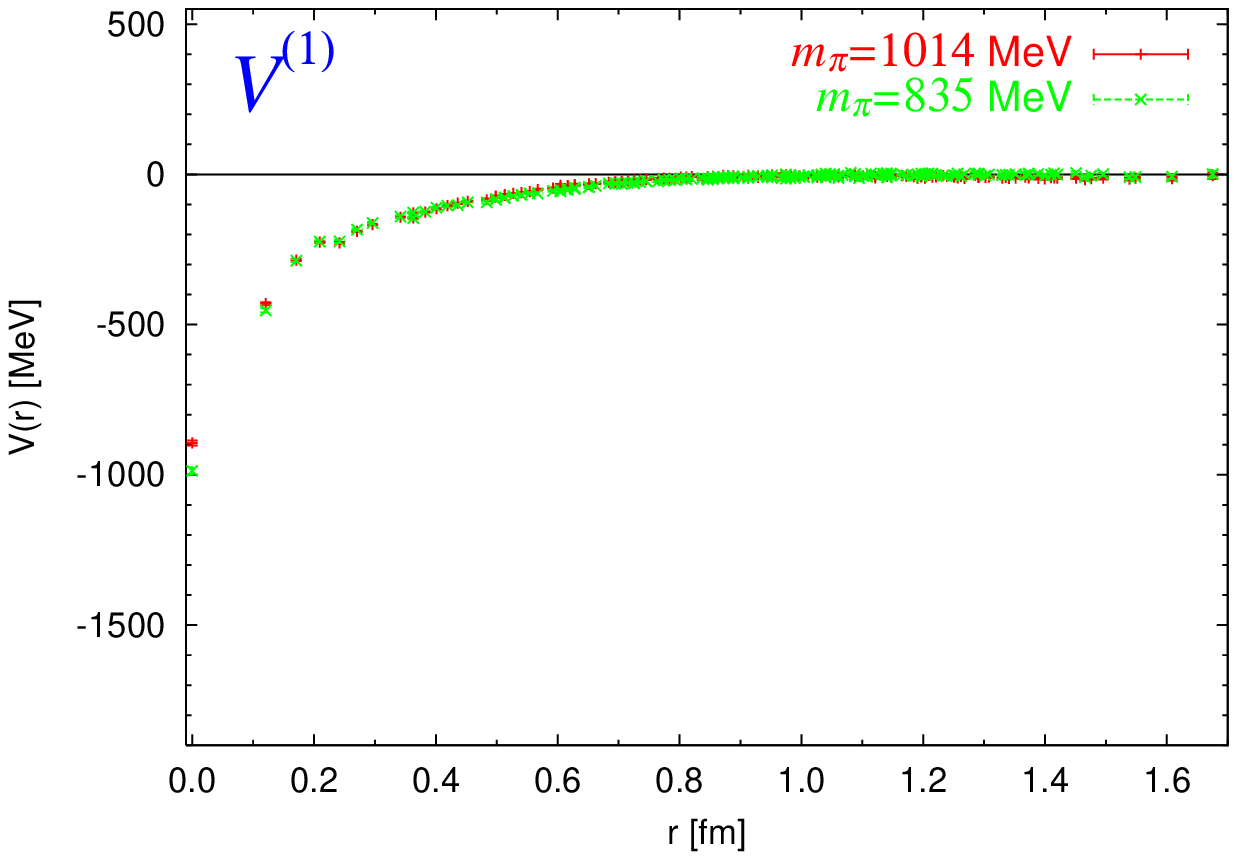}\hfill
 \includegraphics[width=0.49\textwidth]{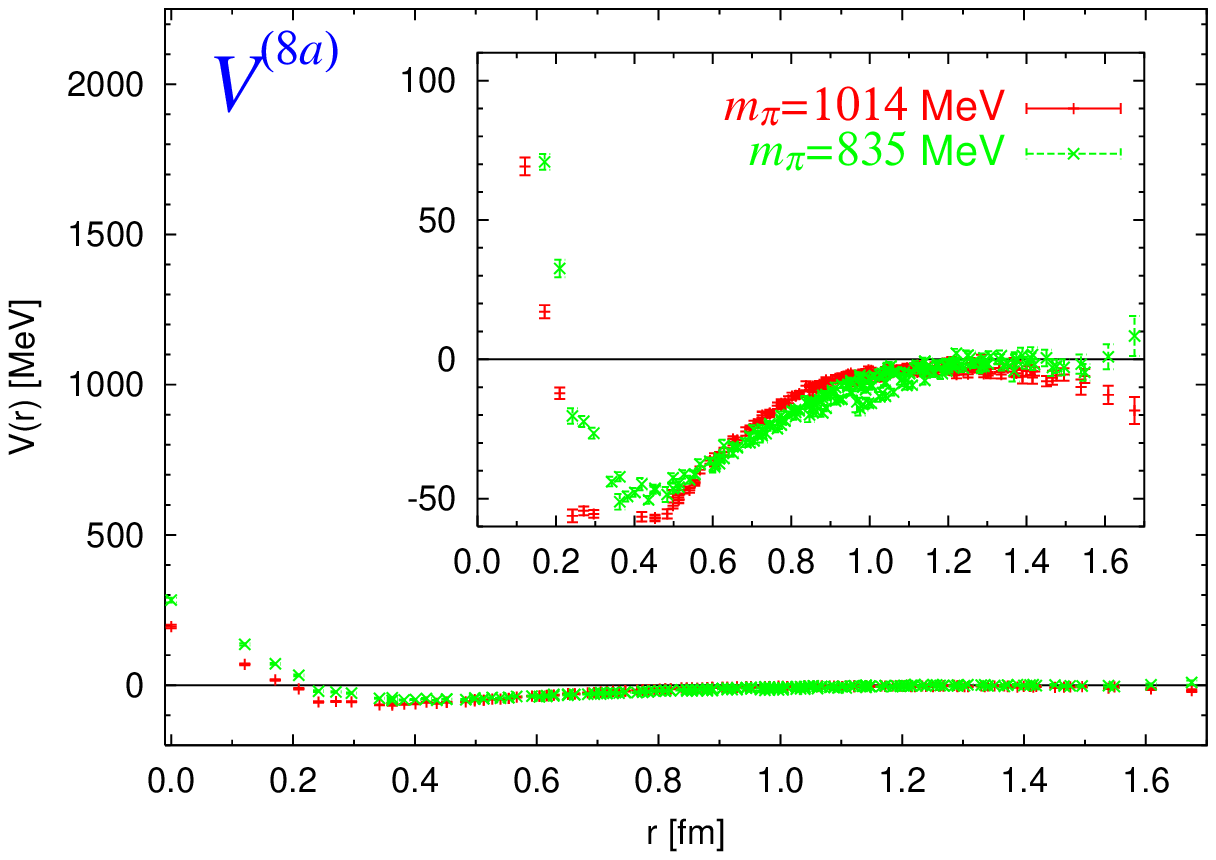}
\caption{\label{fig:pot_2kappa}
  The six independent BB potentials for S-wave in the flavor SU(3) limit, 
  extracted from the lattice QCD simulation at 
  $m_{\pi}=1014$ MeV (red bars) and $m_{\pi}=835$ MeV (green crosses).
 } 
\end{figure}

Fig. \ref{fig:pot_2kappa} shows the resulting six BB potentials
in the flavor basis obtained from the NBS wave functions.
Red bars (green crosses) data correspond to the pion mass 1014 MeV (835 MeV):
Although there is a tendency that the magnitude (range) of the  potentials becomes
larger at short distances (longer at large distances) for lighter quark mass,
the differences are not substantial for the present heavy quark masses. 

The left three panels show potentials in the $^1S_0$ channel,
while the right three panels show effective central potentials in the $^3S_1$ channel.
We have estimated the energies $E^{(\alpha)}$ from the condition
that Eq.(\ref{eqn:vr}) approaches zero at large $r$ 
(the method called ``from $V$" in Ref.\cite{Aoki:2005uf}).  
We find $E^{({\bf 27})}  \simeq -5$ MeV,   $E^{({\bf 8}_s)} \simeq 25$ MeV,
$E^{({\bf 1})}  \simeq -30$ MeV, $E^{({\bf 10}^*)} \simeq -10$ MeV,
$E^{({\bf 10})} \simeq  0$ MeV and $E^{({\bf 8}_a)} \simeq -15$ MeV.
Since the constant term (the second term in the right hand side of Eq.(\ref{eqn:vr})) 
is small compared to the Laplacian term (the first term in the right hand side of Eq.(\ref{eqn:vr})) at short distance,
we have not attempted  to extract a precise value of  $E^{(\alpha)}$ in the present  study.

\begin{table}[t]
 \caption{\label{tbl:pairs}Baryon pairs in an irreducible flavor SU(3) representation,
 where $\{BB'\}$ and $[BB']$ denotes $BB' + B'B$ and $BB' - B'B$, respectively.}
   \begin{center}
   \begin{tabular}{c|l}
   \hline \hline
   flavor multiplet      & baryon pair (isospin)                      \\
   \hline 
   {\bf 27}       & \{$NN$\}(I=1), \{$N\Sigma$\}(I=3/2), \{$\Sigma\Sigma$\}(I=2),  \\
                  & \{$\Sigma\Xi$\}(I=3/2), \{$\Xi\Xi$\}(I=1) \\
   {\bf 8}$_s$      & none                             \\
   ~{\bf 1}       & none                             \\  
  \hline 
  {\bf 10}$^*$    & [$NN$](I=0), [$\Sigma\Xi$](I=3/2)    \\   
  {\bf 10}        & [$N\Sigma$](I=3/2), [$\Xi\Xi$](I=0)\\
  {\bf 8}$_a$     & [$N\Xi$](I=0)       \\
   \hline \hline
 \end{tabular}
 \end{center}
\end{table}

Some of octet-baryon pairs belong exclusively to one irreducible representation, 
as shown in Table \ref{tbl:pairs}.
For example, a symmetric NN belongs to ${\bf 27}$ so that
one can regard $V^{({\bf 27})}$ as a SU(3) limit of  the NN $^1S_0$ potential. 
Similarly,  $V^{({\bf 10}^*)}$, $V^{({\bf 10})}$ and $V^{({\bf 8}_a)}$ 
can be considered as the SU(3) limits of some potentials in the baryon basis,
while $V^{({\bf 8}_s)}$ and $V^{({\bf 1})}$ are always 
superpositions of different potentials in the baryon basis.

Top two panels of Fig. \ref{fig:pot_2kappa} show $V^{({\bf 27})}$ 
and  $V^{({\bf 10}^*)}$, which  corresponds to NN $^1S_0$ potential 
and NN $^3S_1$ potential respectively.
Both have a repulsive core at short distance and an attractive pocket around 0.6 fm.
These qualitative features are consistent with the previous results found in the NN system
in quenched approximation with lighter quark mass\cite{Ishii:2006ec}.
The right middle panel of Fig. \ref{fig:pot_2kappa} shows that $V^{({\bf 10})}$ has
a stronger repulsive core and a weaker attractive pocket than  $V^{({\bf 27},{\bf 10}^*)}$. 
Furthermore $V^{({\bf 8}_s)}$ in the left middle panel of Fig.\ref{fig:pot_2kappa} has
a very strong repulsive core among all  channels,
while $V^{({\bf 8}_a)}$ in the right bottom panel of Fig.\ref{fig:pot_2kappa} has a very weak repulsive core.
In contrast to other cases,  
$V^{({\bf 1})}$  shows attraction for all distances instead of repulsion  at short distance,
as shown in the left bottom panel of Fig. \ref{fig:pot_2kappa}.

Above features are consistent with what has been 
observed in phenomenological quark models\cite{Oka:2000wj}.
In particular, the potential in the ${\bf 8}_s$ channel in quark models
becomes strongly repulsive
at short distance since the six quarks cannot occupy the same orbital state
 due to quark Pauli blocking.  On the other hand,
 the potential in the ${\bf 1}$ channel
does not suffer from the quark Pauli blocking and can become attractive
due to short range gluon exchange. 
 Such an agreement between the lattice data and the 
 phenomenological models suggests that the quark Pauli blocking plays an 
 essential role for the repulsive core in BB systems as originally 
 proposed in Ref.\cite{Otsuki:1965yk}.

\subsection{BB potentials in baryon basis}

The BB potentials in the baryon basis can be obtained by  
an unitary rotation of those in the flavor basis.
Its explicit form reads, 
\begin{equation}
  V_{ij}(r) = \sum_{\alpha} U_{i\alpha} \,  V^{(\alpha)}(r) \, U^{\dagger}_{\alpha j}
\label{eq:rotation}
\end{equation}
where $U$ is an unitary matrix which rotates the flavor basis $\{\brav{\alpha}|\}$ 
to baryon basis $\{\brav i|\}$, {\it i.e.} $\brav i| = \sum_{\alpha} U_{i\alpha} \brav{\alpha}|$.
The explicit forms of the unitary matrix $U$ in terms of the 
CG coefficients are given in Appendix \ref{app:cgcoeff}.

In Fig. \ref{fig:pot_lamlam}, as characteristic examples,
we show the potentials for S=$-$2, I=0 sector in the $^1S_0$ channel at $m_\pi = 835$ MeV. 
To obtain $V_{ij}(r)$,  we first fit the potentials in the flavor basis
by the following form with five parameters $b_1 \sim b_5$,
\begin{equation}
V(r) = b_1 e^{-b_2\,r^2} + b_3(1 - e^{-b_4\,r^2})\left( \frac{e^{-b_5\,r}}{r} \right)^2 .
\end{equation}
We then use the right hand side of  Eq.(\ref{eq:rotation})
to obtain the potentials in the baryon basis.
The left panel of  Fig.\ref{fig:pot_lamlam} shows the diagonal part of the potentials.
The strong repulsion in the ${\bf 8}_s$  channel is reflected most in the $\Sigma\Sigma$(I=0) potential
due to its largest CG coefficient among three channels.
The strong attraction in the ${\bf 1}$ channel is reflected most in the $N\Xi$(I=0) potential
due to its largest CG coefficient. 
Nevertheless, all three diagonal potentials have repulsive core originating from the ${\bf 8}_s$ component.
 The right panel of  Fig. \ref{fig:pot_lamlam} shows the off-diagonal part of the potentials
 which are comparable in magnitude to the diagonal ones.
 Since the off-diagonal parts are not negligible in the baryon basis,
 full coupled channel analysis is necessary to study observables.
 Similar situation  holds in (2+1)-flavors: The flavor basis with approximately
 diagonal potentials are useful for obtaining essential features of the BB interactions,
 while the baryon basis with substantial magnitude of the off-diagonal potentials are
 necessary for practical applications.

\begin{figure}[t]
 \includegraphics[width=0.49\textwidth]{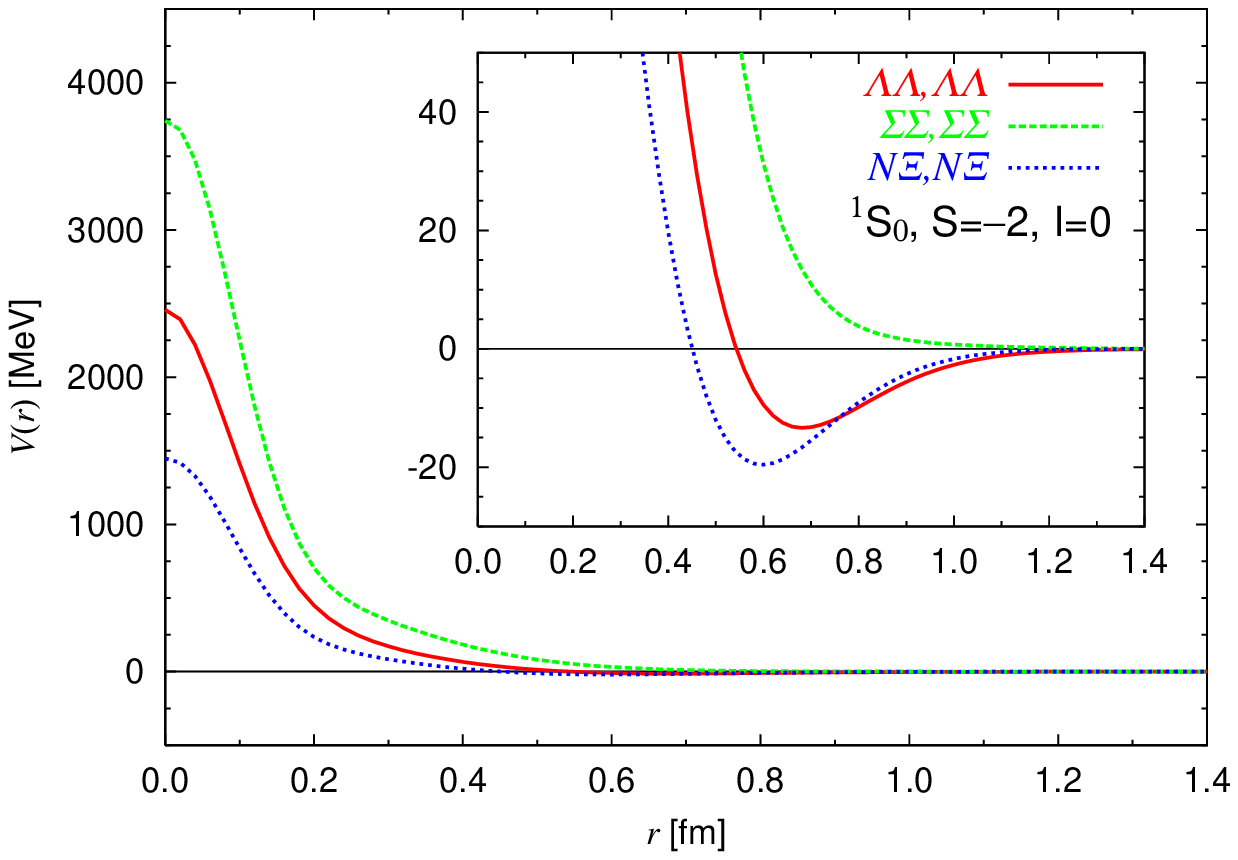}\hfill
 \includegraphics[width=0.49\textwidth]{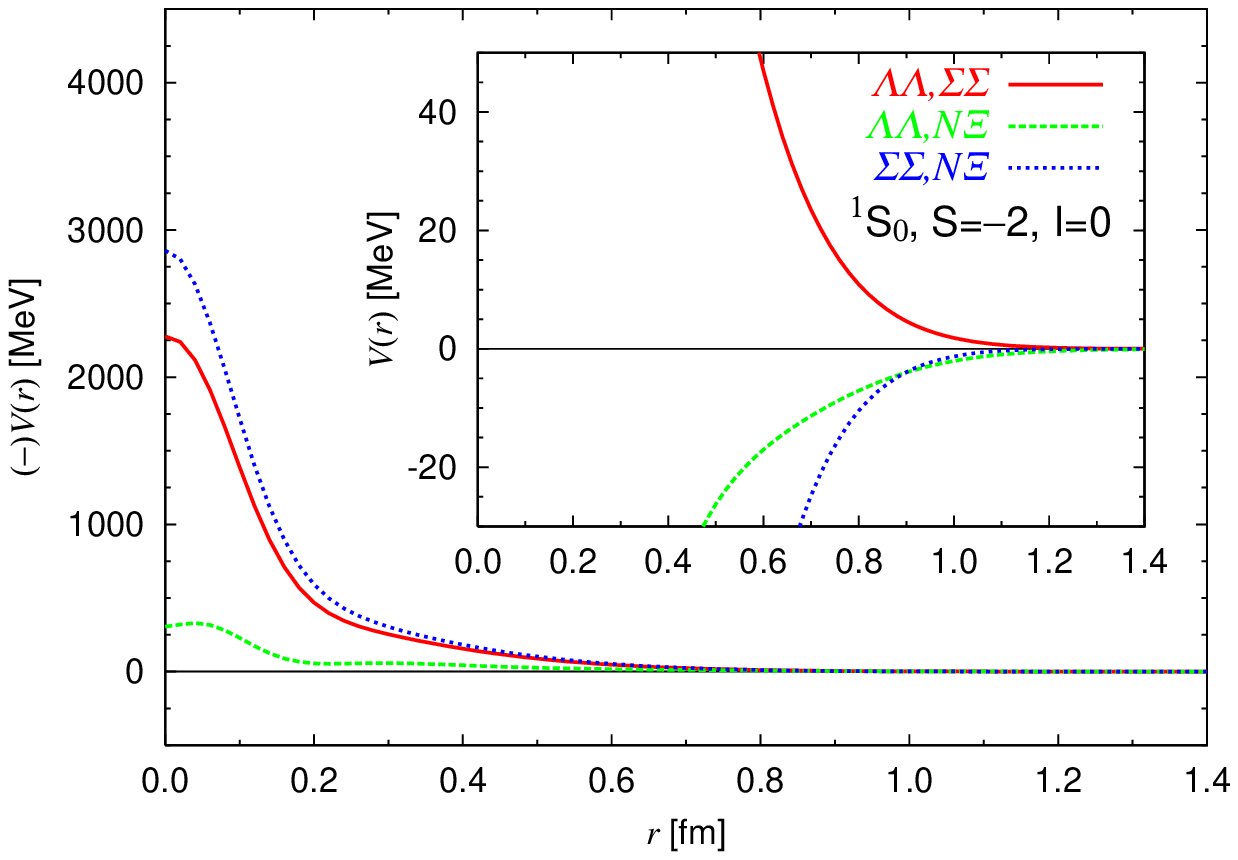}
 \caption{\label{fig:pot_lamlam}
  BB potentials in baryon basis for S=$-$2, I=0, $^1S_0$ sector. 
  Three diagonal(off-diagonal) potentials are shown in left(right) panel.
  Phase of off-diagonal ones in the right panel are arranged in zoom-out plot. 
  Their true signs are shown in the insertion.
 }
\end{figure}

\begin{figure}[tp]
 \includegraphics[width=0.49\textwidth]{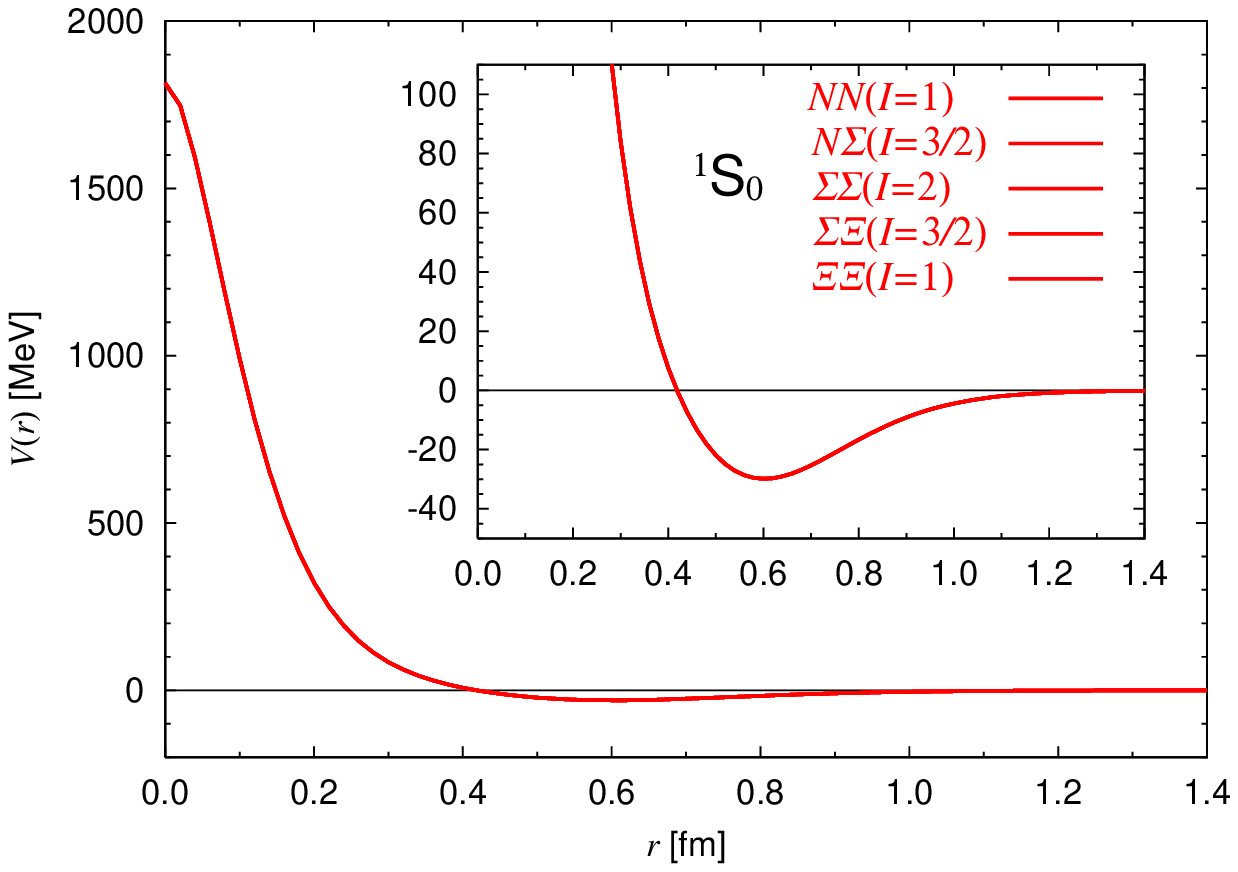}\hfill
 \includegraphics[width=0.49\textwidth]{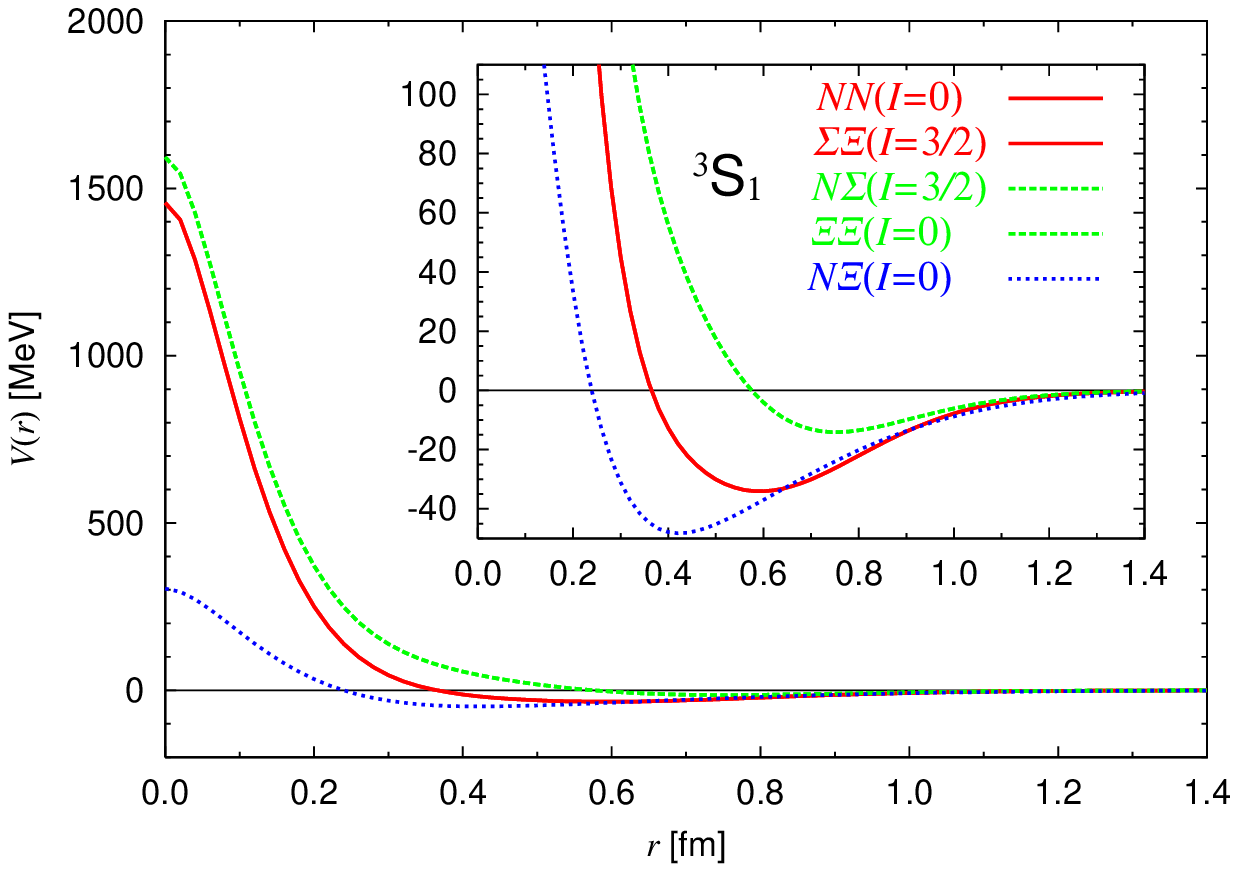}
 \includegraphics[width=0.49\textwidth]{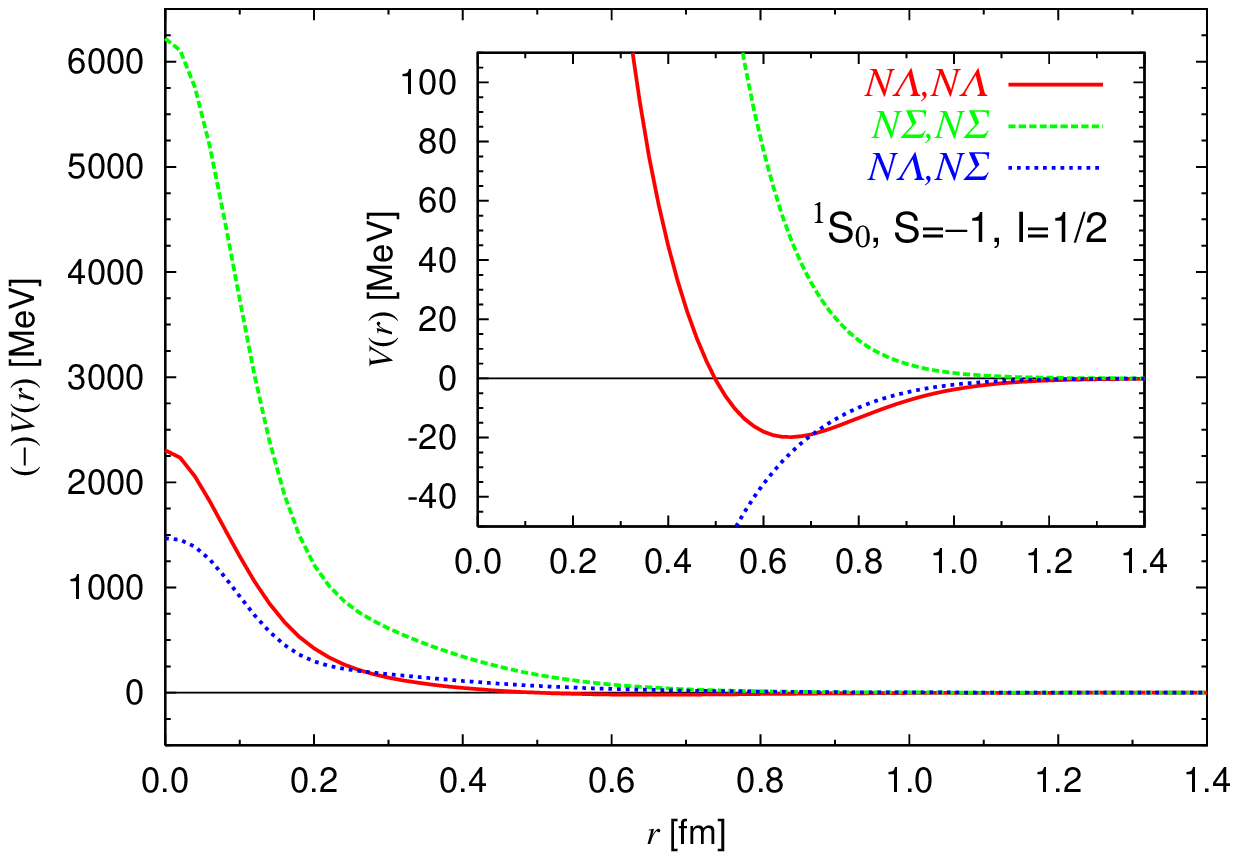}\hfill
 \includegraphics[width=0.49\textwidth]{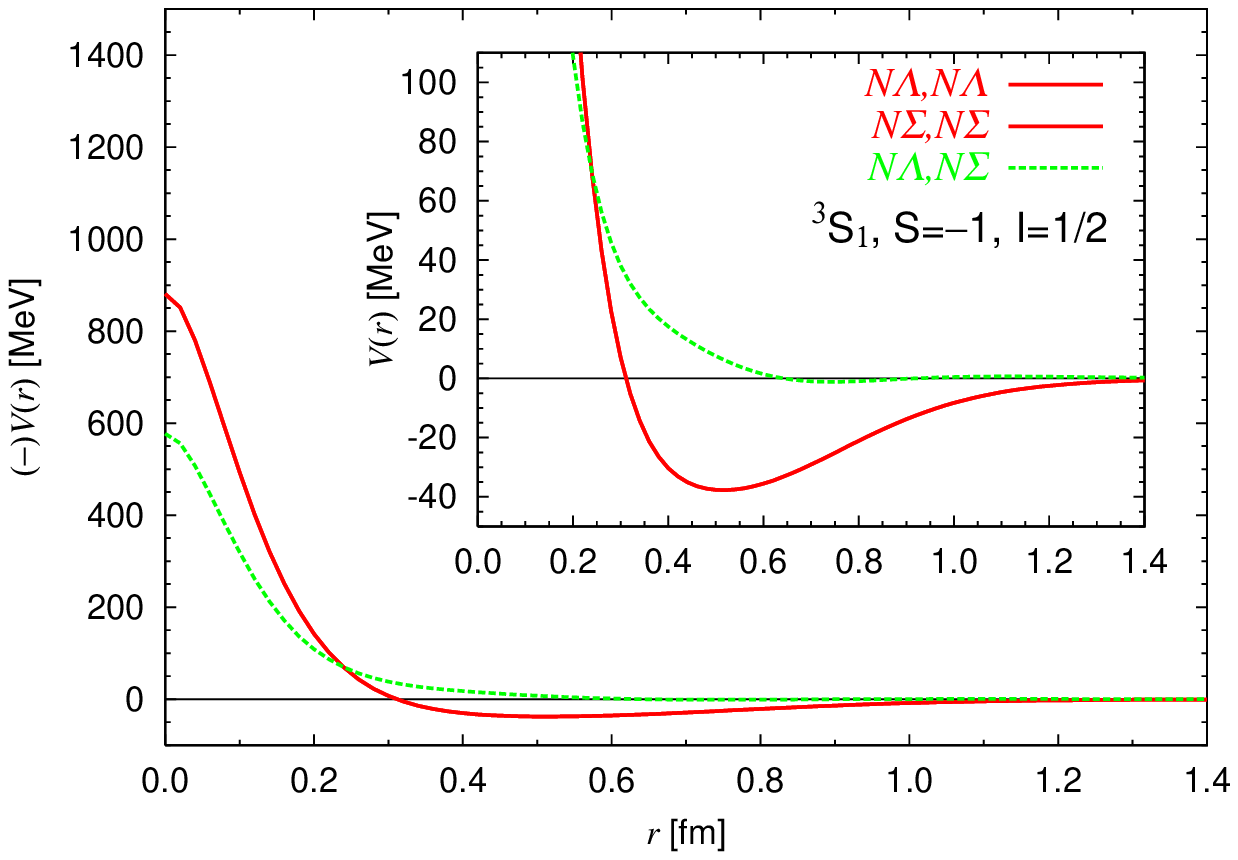}
 \includegraphics[width=0.49\textwidth]{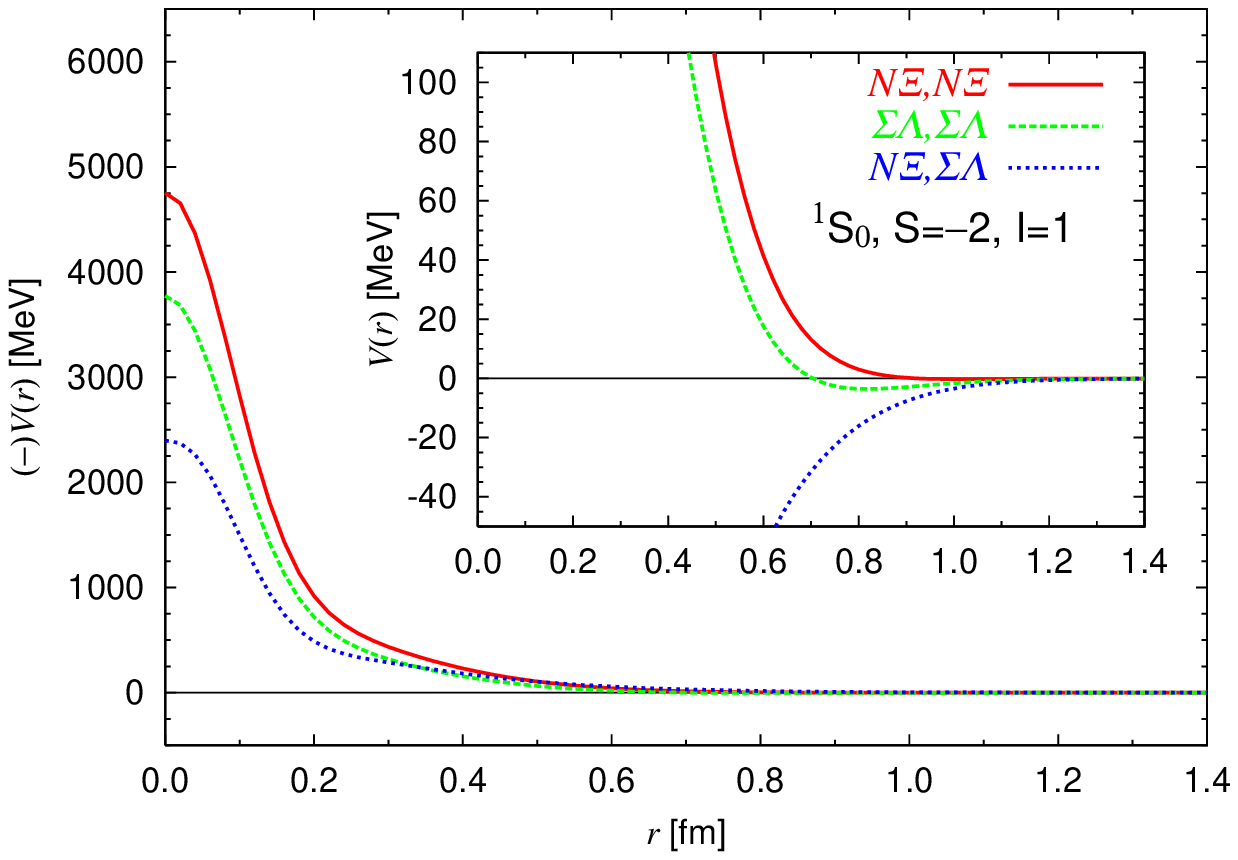}\hfill
 \includegraphics[width=0.49\textwidth]{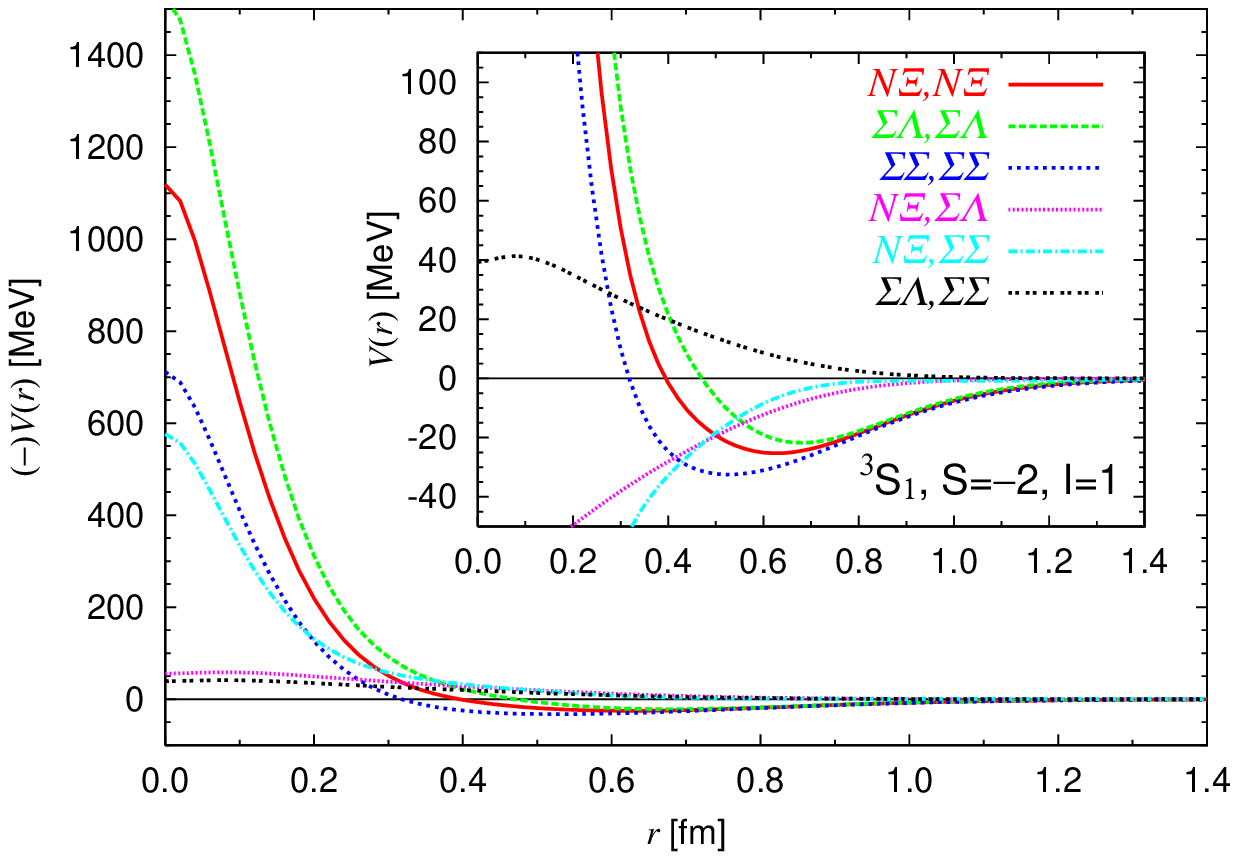}
 \includegraphics[width=0.49\textwidth]{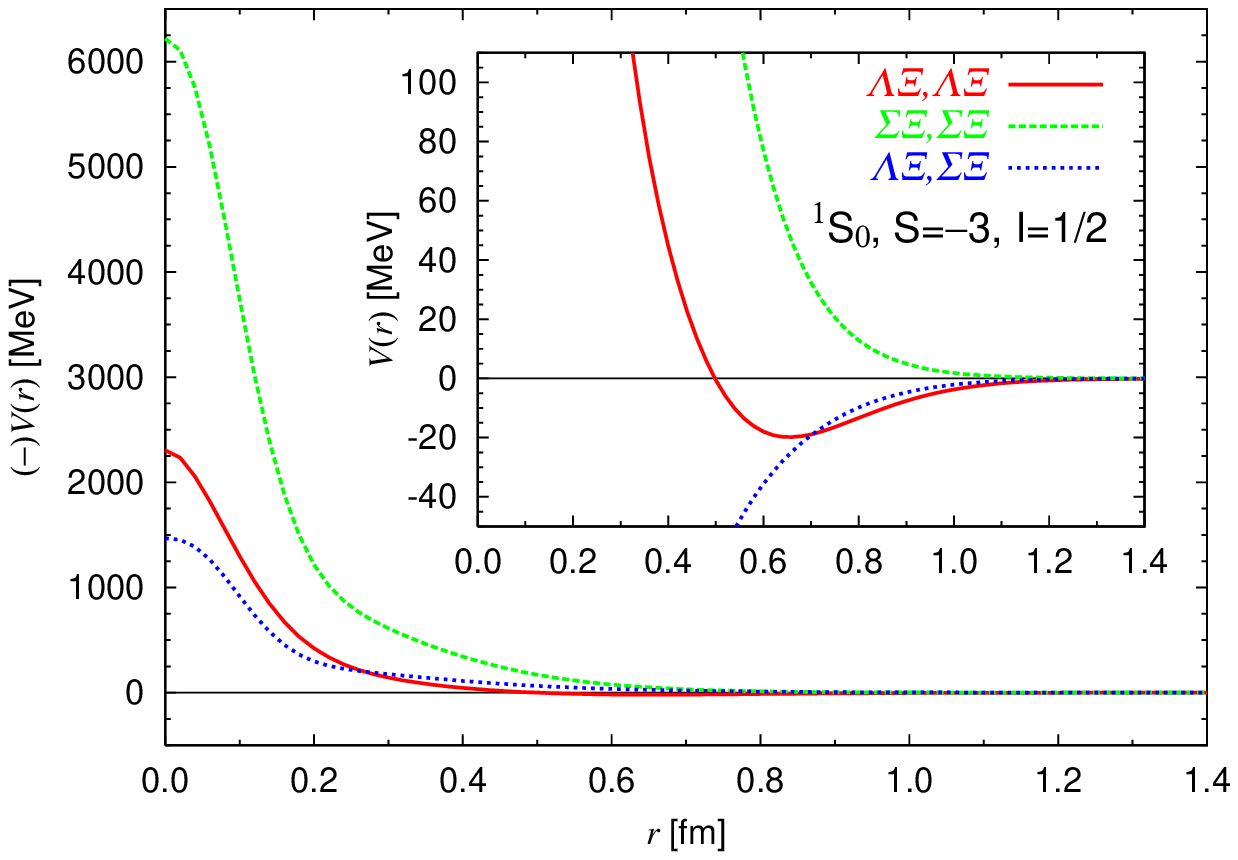}\hfill
 \includegraphics[width=0.49\textwidth]{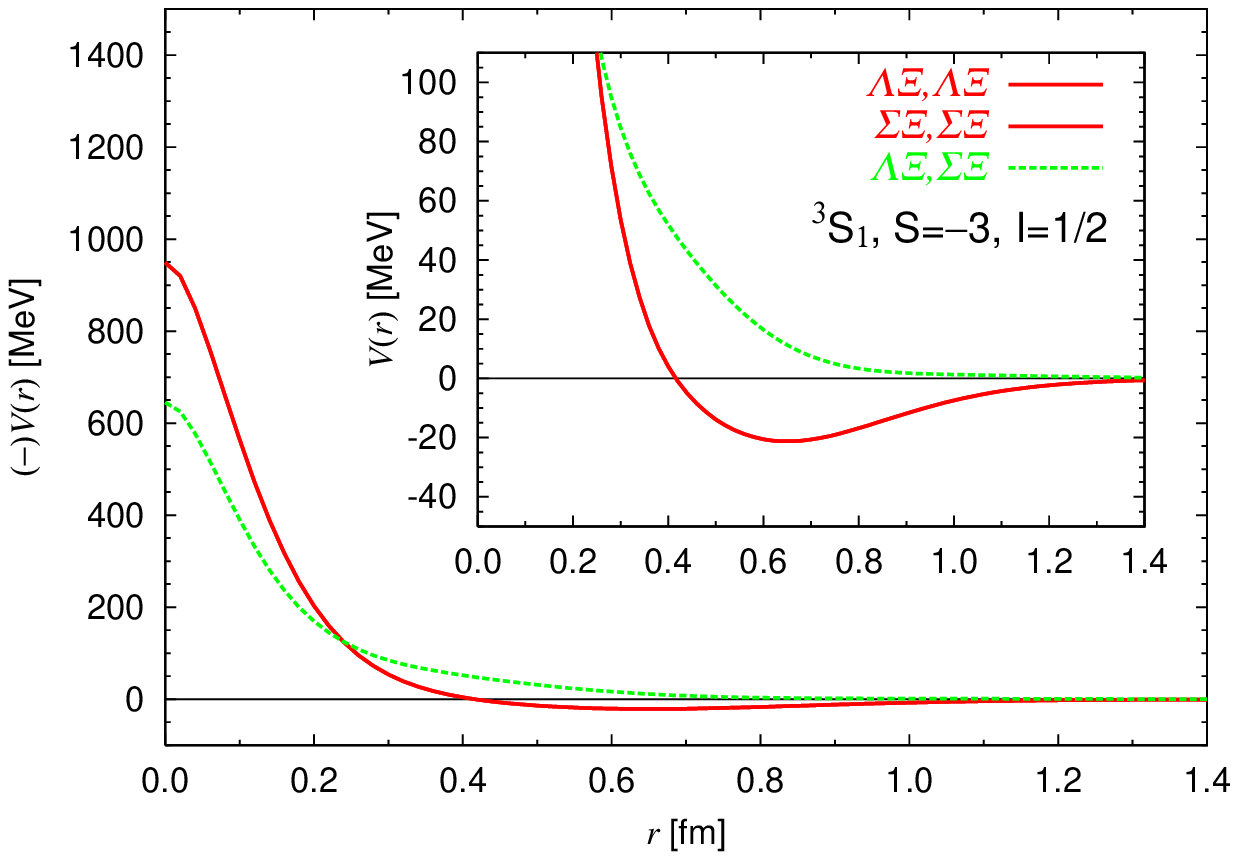}
 \caption{\label{fig:pot_particle}
  BB potentials in baryon basis other than those shown in Fig.\ref{fig:pot_lamlam}. 
  See the caption of Fig.\ref{fig:pot_lamlam}.
 }
\end{figure}

Other potentials in baryon basis are given in Fig. \ref{fig:pot_particle}.
Since the ${\bf 8}_s$ state does not couple to the $^3S_1$ channel,
the repulsive cores in the $^3S_1$ channel (the right panels of Fig. \ref{fig:pot_particle}) 
are relatively small.  
The off-diagonal potentials are not generally small: 
In the right second panel of Fig.\ref{fig:pot_particle}, for example,
the  $N\Lambda$-$N\Sigma$ potential in the $^3S_1$ channel 
is comparable in magnitude at short distances with the 
diagonal $N\Lambda$-$N\Lambda$ and $N\Sigma$-$N\Sigma$ potentials.
Although all quark masses of 3 flavors are degenerate and rather heavy in our simulations,
these coupled channel potentials in the baryon basis may give 
useful hints for the behavior of hyperons ($\Lambda$, $\Sigma$ and $\Xi$)
in hypernuclei and in neutron stars \cite{Hashimoto:2006aw,SchaffnerBielich:2010am}.

\section{Summary and outlook}

We have performed  3-flavor full QCD simulations
to study the general features of the BB interaction in the flavor SU(3) limit.
From the NBS wave function measured on the lattice, 
we extracted all six independent potentials in the S-wave at
 the leading order of the derivative expansion;
$V^{({\bf 27})}$, $V^{({\bf 8}_s)}$, $V^{({\bf 1})}$,
$V^{({\bf 10}^*)}$, $V^{({\bf 10})}$ and $V^{({\bf 8}_a)}$
labeled by flavor irreducible representation.
We have found strong flavor dependence of the BB interactions. 
In particular, $V^{({\bf 8}_s)}$ has a very strong repulsive core at short distance, 
while $V^{({\bf 1})}$ is attractive in all distances. 
These features are qualitatively consistent with the Pauli blocking effect
among quarks previously studied in phenomenological quark models. 
Recently, the existence (absence)  of the repulsive core
in the $K^+$-$P$ potential \cite{Ikeda:2010sg}  (in the $\bar{c}c$-$N$ potential \cite{Kawanai:2010cq})
has been observed in the same method as adopted in this paper. These results 
further support the relevance of the Pauli blocking effect in the quark level for hadron interactions.  

The BB potentials in the baryon basis are reconstructed from those in the flavor basis.
We have found that the potentials in $^3S_1$ channels have
weaker repulsive core than the  $^1S_0$ channels,
since the ${\bf 8}_s$ state, which has the strongest repulsive core, does not mix with the former.
We note that the off-diagonal potentials are not generally small compared to the diagonal ones.
These information together with the future lattice simulations at
non-degenerate quark masses in larger volumes will give 
useful hints for the structure of hypernuclei and neutron stars.
 
The flavor singlet channel has attraction for all distances in our simulation. 
The present data are not sufficient to make a definite
conclusion on the $H$-dibaryon,
because of single small lattice volume and a large degenerate quark mass.
Nevertheless, by using the potential $V^{({\bf 1})}(r)$ and solving the 
Schr\"{o}dinger equation Eq.(\ref{eq:shroedinger}) with $E^{({\bf 1})}=-30$ MeV, 
we find a shallow bound state, 
which suggests a possibility of a bound $H$-dibaryon with baryon components
$\Lambda\Lambda:\Sigma\Sigma:\Xi N =-\sqrt{1}:\sqrt{3}:\sqrt{4}$
in the SU(3) symmetric world at large quark masses. 
In the real world, the baryon mass ordering  in S=$-$2 and I=0 sector
becomes $\Lambda\Lambda < \Xi N < \Sigma\Sigma$ due to flavor-SU(3) breaking.
In our trial analysis of solving the Schr\"{o}dinger equation by using
the potentials obtained in this paper together with
small baryon mass differences from a 2+1 flavor lattice
QCD simulation\footnote{
With hopping parameters $\kappa_{ud}=0.13760$ and $\kappa_{s}=0.13710$
at $\beta=1.83$, baryon masses are obtained
$M_{N}=1797$ MeV, $M_{\Lambda}=1827$ MeV, $M_{\Sigma}=1833$ MeV, $M_{\Xi}=1860$ MeV.}, 
a resonance state is found at an energy between $\Lambda\Lambda$ mass and $\Xi N$ mass.
If such a resonance  exists in nature,
it may explain the enhancement just above the $\Lambda \Lambda$ threshold recently reported
in the KEK experiment \cite{Yoon:2007aq}.
However, further investigations in both theory and experiment
are necessarily to draw a definite conclusion.

\section*{Acknowledgements}
The authors thank the CP-PACS and JLQCD Collaborations for their configurations,
the Columbia Physics System \cite{CPS} for their lattice QCD simulation code,
and M. Oka, K. Yazaki, A. Ohnishi, K. Imai, C. Nakamoto, Y. Fujiwara,
T. Kawanai and S. Sasaki for helpful discussions and informations.
This research is supported in part by 
Grant-in-Aid for Scientific Research on Innovative Areas(No.2004:20105001, 20105003)
and the Large Scale Simulation Program No.09-23(FY2009) of High Energy Accelerator Research Organization(KEK). 
S. A. and T. I. are supported in part by the Grant-in-Aid of the Ministry of Education,
Science and Technology, Sports and Culture(No.20340047).
N. I. is supported in part by Grant-in-Aid of MEXT(No.22540268)
and a Grand-in-Aid for Specially Promoted Research(13002001).
T. D. is supported in part by Grant-in-Aid for JSPS Fellows 21$\cdot$5985.
H. N. is supported in part by the MEXT Grant-in-Aid,
Scientific Research on Innovative Areas(No.21105515).
\appendix
\section{\label{app:source}The irreducible BB operators}
In this Appendix, flavor irreducible BB operators used in this study are given.
The  composite operators for octet baryons  are
\begin{eqnarray}
p_{\alpha}(x) &=& +\epsilon_{c_1,c_2,c_3}\,(C\gamma_5)_{\beta_1,\beta_2}\,\delta_{\beta_3,\alpha}\,u(\xi_1)d(\xi_2)u(\xi_3)
\\
n_{\alpha}(x) &=& +\epsilon_{c_1,c_2,c_3}\,(C\gamma_5)_{\beta_1,\beta_2}\,\delta_{\beta_3,\alpha}\,u(\xi_1)d(\xi_2)d(\xi_3)
\\
\Sigma^+_{\alpha}(x) &=& -\epsilon_{c_1,c_2,c_3}\,(C\gamma_5)_{\beta_1,\beta_2}\,\delta_{\beta_3,\alpha}\,u(\xi_1)s(\xi_2)u(\xi_3)
\\
\Sigma^0_{\alpha}(x) &=& -\epsilon_{c_1,c_2,c_3}\,(C\gamma_5)_{\beta_1,\beta_2}\,\delta_{\beta_3,\alpha}\,
                \sqrt{\frac12}\left[ d(\xi_1)s(\xi_2)u(\xi_3) + u(\xi_1)s(\xi_2)d(\xi_3) \right]
\\
\Sigma^-_{\alpha}(x) &=& -\epsilon_{c_1,c_2,c_3}\,(C\gamma_5)_{\beta_1,\beta_2}\,\delta_{\beta_3,\alpha}\,d(\xi_1)s(\xi_2)d(\xi_3)
\\
\Xi^0_{\alpha}(x) &=& +\epsilon_{c_1,c_2,c_3}\,(C\gamma_5)_{\beta_1,\beta_2}\,\delta_{\beta_3,\alpha}\,s(\xi_1)u(\xi_2)s(\xi_3)
\\
\Xi^-_{\alpha}(x) &=& +\epsilon_{c_1,c_2,c_3}\,(C\gamma_5)_{\beta_1,\beta_2}\,\delta_{\beta_3,\alpha}\,s(\xi_1)d(\xi_2)s(\xi_3)
\\
\Lambda_{\alpha}(x) &=& -\epsilon_{c_1,c_2,c_3}\,(C\gamma_5)_{\beta_1,\beta_2}\,\delta_{\beta_3,\alpha} \times \nonumber \\
  & & \qquad 
 \sqrt{\frac16}\left[ d(\xi_1)s(\xi_2)u(\xi_3) + s(\xi_1)u(\xi_2)d(\xi_3) - 2 u(\xi_1)d(\xi_2)s(\xi_3) \right]
\end{eqnarray}
with a notation $\xi_i = \{c_i,\beta_i,x\}$. 
We follow the phase convention in Ref. \cite{deSwart:1963gc} .
 In this convention, the two-baryon operator which belongs to the definite flavor representation
can be constructed with the Clebsch-Gordan coefficient of SU(3) as
\begin{eqnarray}
{BB}^{({\bf 27})} &=& +\sqrt{\frac{27}{40}}{\Lambda}\,{\Lambda}
                         -\sqrt{\frac{1}{40} }{\Sigma}\,{\Sigma}
                         +\sqrt{\frac{12}{40}}{N}\,{\Xi}
\\
{BB}^{({\bf 8}_s)} &=& -\sqrt{\frac15}{\Lambda}\,{\Lambda}
                         -\sqrt{\frac35}{\Sigma}\,{\Sigma}
                         +\sqrt{\frac15}{N}\,{\Xi}
\\
{BB}^{({\bf 1})} &=& -\sqrt{\frac18}{\Lambda}\,{\Lambda}
                        +\sqrt{\frac38}{\Sigma}\,{\Sigma}
                        +\sqrt{\frac48}{N}\,{\Xi}
\\
{BB}^{({\bf 10}^*)} &=& +\sqrt{\frac12}{p}\,{n}
                           -\sqrt{\frac12}{n}\,{p}
\\
{BB}^{({\bf 10})}  &=& +\sqrt{\frac12}{p}\,{\Sigma^+}
                           -\sqrt{\frac12}{\Sigma^+}\,{p}
\\
{BB}^{({\bf 8}_a)}  &=& +\sqrt{\frac14}{p}\,{\Xi^-}
                           -\sqrt{\frac14}{\Xi^-}\,{p}
                           -\sqrt{\frac14}{n}\,{\Xi^0}
                           +\sqrt{\frac14}{\Xi^0}\,{n}
\end{eqnarray}
where ${\Sigma}\,{\Sigma}$ and ${N}\,{\Xi}$ represents
\begin{eqnarray}
{\Sigma}{\Sigma} &=& +\sqrt{\frac13}{\Sigma^+}\,{\Sigma^-}
                     -\sqrt{\frac13}{\Sigma^0}\,{\Sigma^0}
                     +\sqrt{\frac13}{\Sigma^-}\,{\Sigma^+}
\\
{N}{\Xi} &=& +\sqrt{\frac14}{p}\,{\Xi^-}
                     +\sqrt{\frac14}{\Xi^-}\,{p}
                     -\sqrt{\frac14}{n}\,{\Xi^0}
                     -\sqrt{\frac14}{\Xi^0}\,{n} ~.
\end{eqnarray}

\section{\label{app:cgcoeff}Relation between the flavor basis and baryon basis}
Unitary matrices $U$ 
which rotates the flavor basis $\{\brav{\alpha}|\}$ to baryon basis $\{\brav i|\}$,
are given as follows.
\begin{enumerate}
\item S=$-$1, I=1/2, $^1S_0$ sector.
\begin{eqnarray}
 \left(
  \begin{array}{l}
   \brav N\Lambda| \\
   \brav N\Sigma|
  \end{array}
  \right)
=
   \left(
    \begin{array}{rr}
       \sqrt{\frac{ 9}{10}} &  -\sqrt{\frac{ 1}{10}} \\
       \sqrt{\frac{ 1}{10}} &   \sqrt{\frac{ 9}{10}} 
    \end{array}
   \right)
   \left(
    \begin{array}{l}
     \brav{\bf 27}| \\
     \brav{\bf 8}_s|
    \end{array}
   \right)
\end{eqnarray}

\item S=$-$1, I=1/2, $^3S_1$ sector.
\begin{eqnarray}
\hfill
 \left(
  \begin{array}{l}
   \brav N\Lambda| \\
   \brav N\Sigma|
  \end{array}
  \right)
=
   \left(
    \begin{array}{rr}
       \sqrt{\frac{ 1}{2}} &  -\sqrt{\frac{ 1}{2}} \\
       \sqrt{\frac{ 1}{2}} &   \sqrt{\frac{ 1}{2}} 
    \end{array}
   \right)
   \left(
    \begin{array}{l}
     \brav{\bf 10}^*| \\
     \brav{\bf 8}_a|
    \end{array}
   \right)
\end{eqnarray}

\item S=$-$2, I=0, $^1S_0$ sector.
\begin{equation}
 \left(
  \begin{array}{l}
   \brav \Lambda\Lambda| \\
   \brav \Sigma\Sigma| \\
   \brav N\Xi|
  \end{array}
  \right)
=
   \left(
    \begin{array}{rrr}
       \sqrt{\frac{27}{40}} &  -\sqrt{\frac{ 8}{40}} & -\sqrt{\frac{ 5}{40}}  \\
      -\sqrt{\frac{ 1}{40}} &  -\sqrt{\frac{24}{40}} &  \sqrt{\frac{15}{40}}  \\
       \sqrt{\frac{12}{40}} &   \sqrt{\frac{ 8}{40}} &  \sqrt{\frac{20}{40}}
    \end{array}
   \right)
   \left(
    \begin{array}{l}
     \brav{\bf 27}| \\
     \brav{\bf 8}_s| \\
     \brav{\bf 1}|
    \end{array}
   \right)
\end{equation}

\item S=$-$2, I=1, $^1S_0$ sector.
\begin{eqnarray}
 \left(
  \begin{array}{l}
   \brav N\Xi| \\
   \brav \Sigma\Lambda|
  \end{array}
  \right)
=
   \left(
    \begin{array}{rr}
       \sqrt{\frac{ 2}{5}} &  -\sqrt{\frac{ 3}{5}} \\
       \sqrt{\frac{ 3}{5}} &   \sqrt{\frac{ 2}{5}} 
    \end{array}
   \right)
   \left(
    \begin{array}{l}
     \brav{\bf 27}| \\
     \brav{\bf 8}_s|
    \end{array}
   \right)
\end{eqnarray}

\item S=$-$2, I=1, $^3S_1$ sector.
\begin{eqnarray}
 \left(
  \begin{array}{l}
   \brav N\Xi| \\
   \brav \Sigma\Lambda| \\
   \brav \Sigma\Sigma|
  \end{array}
  \right)
=
   \left(
    \begin{array}{rrr}
      -\sqrt{\frac{ 1}{3}} &  -\sqrt{\frac{ 1}{3}} & \sqrt{\frac{ 1}{3}} \\
      -\sqrt{\frac{ 1}{2}} &   \sqrt{\frac{ 1}{2}} & 0                   \\
       \sqrt{\frac{ 1}{6}} &   \sqrt{\frac{ 1}{6}} & \sqrt{\frac{ 4}{6}} 
    \end{array}
   \right)
   \left(
    \begin{array}{l}
     \brav{\bf 10}^*| \\
     \brav{\bf 10}| \\
     \brav{\bf 8}_a|
    \end{array}
   \right)
\end{eqnarray}

\item S=$-$3, I=1/2, $^1S_0$ sector.
\begin{eqnarray}
 \left(
  \begin{array}{l}
   \brav \Lambda\Xi| \\
   \brav \Sigma\Xi|
  \end{array}
  \right)
=
   \left(
    \begin{array}{rr}
      \sqrt{\frac{ 9}{10}} &  -\sqrt{\frac{ 1}{10}}  \\
      \sqrt{\frac{ 1}{10}} &   \sqrt{\frac{ 9}{10}} 
    \end{array}
   \right)
   \left(
    \begin{array}{l}
     \brav{\bf 27}| \\
     \brav{\bf 8}_s|
    \end{array}
   \right)
\end{eqnarray}

\item S=$-$3, I=1/2, $^3S_1$ sector.
\begin{eqnarray}
 \left(
  \begin{array}{l}
   \brav \Lambda\Xi| \\
   \brav \Sigma\Xi|
  \end{array}
  \right)
=
   \left(
    \begin{array}{rr}
      \sqrt{\frac{ 1}{2}} &  -\sqrt{\frac{ 1}{2}}  \\
      \sqrt{\frac{ 1}{2}} &   \sqrt{\frac{ 1}{2}} 
    \end{array}
   \right)
   \left(
    \begin{array}{l}
     \brav{\bf 10}| \\
     \brav{\bf 8}_a|
    \end{array}
   \right)
\end{eqnarray}
\end{enumerate}

\end{document}